\documentclass{article}

\PassOptionsToPackage{numbers, compress}{natbib}
\usepackage[preprint]{neurips_2019}

\usepackage[utf8]{inputenc} % allow utf-8 input
\usepackage[T1]{fontenc}    % use 8-bit T1 fonts
\usepackage{hyperref}       % hyperlinks
\usepackage{url}            % simple URL typesetting
\usepackage{booktabs}       % professional-quality tables
\usepackage{amsfonts}       % blackboard math symbols
\usepackage{nicefrac}       % compact symbols for 1/2, etc.
\usepackage{microtype}      % microtypography
\usepackage{graphicx}
\usepackage{algorithm}
\usepackage[noend]{algpseudocode}

\title{DynWalks: Global Topology and Recent Changes Awareness Dynamic Network Embedding}

\author{
  Chengbin Hou\thanks{This work is done while Chengbin Hou is studying at School of Computer Science, The University of Birmingham, UK. Chengbin Hou and Han Zhang contributed equally.} \\
  Department of Computer Science and Engineering \\
  Southern University of Science and Technology, China \\
  \texttt{chengbin.hou10@foxmail.com} \\
  \And
  Han Zhang \\
  School of Computer Science \\
  University of Birmingham, UK \\
  \texttt{hxz325@cs.bham.ac.uk} \\
  \And
  Ke Tang \\
  Department of Computer Science and Engineering \\
  Southern University of Science and Technology, China \\
  \texttt{tangk3@sustech.edu.cn} \\
  \And
  Shan He \\
  School of Computer Science \\
  University of Birmingham, UK \\
  \texttt{s.he@cs.bham.ac.uk} \\
}

\begin{document}

\maketitle

\begin{abstract}
Learning topological representation of a network in dynamic environments has recently attracted considerable attention due to the time-evolving nature of many real-world networks i.e. nodes/links might be added/removed as time goes on. Dynamic network embedding aims to learn low dimensional embeddings for unseen and seen nodes by using any currently available snapshots of a dynamic network. For seen nodes, the existing methods either treat them equally important or focus on the $k$ most affected nodes at each time step. However, the former solution is time-consuming, and the later solution that relies on incoming changes may lose the global topology---an important feature for downstream tasks. To address these challenges, we propose a dynamic network embedding method called DynWalks, which includes two key components: 1) An online network embedding framework that can dynamically and efficiently learn embeddings based on the selected nodes; 2) A novel online node selecting scheme that offers the flexible choices to balance global topology and recent changes, as well as to fulfill the real-time constraint if needed. The empirical studies on six real-world dynamic networks under three different slicing ways show that DynWalks significantly outperforms the state-of-the-art methods in graph reconstruction tasks, and obtains comparable results in link prediction tasks. Furthermore, the wall-clock time and complexity analysis demonstrate its excellent time and space efficiency. The source code of DynWalks is available at \url{https://github.com/houchengbin/DynWalks}.
\end{abstract}

% DNC现实网络。。。。DNC-30D 不行，原因是太稀疏，引用自己文章；用反例来说明我们是全局保留，但是DNC是局部。。。所以我们做不好。。。limitation，未来可以用attribute弥补。。。[cite自己]。。。。 另外，recent change机制也起作用了，因为CGR比GR我们和Triad比，自己进步了！

\section{Introduction}
The interactions or connectivities between entities of a real-world complex system can be naturally represented as a network (or graph) e.g. social networks, biological networks, and sensor networks. Learning topological representation of a network, especially low dimensional node embeddings that encode network topology therein so as to facilitate downstream tasks, has received a great success in the past few years \cite{cui2018survey, hamilton2017representation}. 
For example, DeepWalk \cite{perozzi2014deepwalk} conducts truncated random walks on each node to generate node sequences which are then fed into the Skip-Gram neural network model \cite{mikolov2013distributed} for training node embeddings; Node2Vec \cite{grover2016node2vec} extends DeepWalk by using more flexible truncated random walks to explore network topology; LINE \cite{tang2015line} explicitly includes both the first-order and second-order proximity of each nodes in the KL-divergence objective function; HOPE \cite{ou2016asymmetric} considers the high-order proximity and obtains node embeddings by seeking the largest $k$ singular values of the generalized SVD problem; and etc. 
Note that these methods are designed for static networks.
% Besides, SDNE \cite{wang2016structural} uses the static de-noise auto-enconder and GCN \cite{kipf2017semi} define a graph convolutional network to aggregate node attributes.

However, many real-world networks are dynamic by nature i.e. nodes/links might be added/removed as time goes on. For example, in a wireless sensor network, devices will regularly or accidentally connect to or disconnect from the routers; in a social network, new users and new friendships between existing users will occur as time goes on. The aforementioned static network embedding methods cannot properly deal with dynamic networks, since they cannot update embeddings for the seen nodes based on incoming changes, and may require retraining embeddings for all nodes from scratch. 

Dynamic network embedding, which aims to learn low dimensional embeddings for unseen and seen nodes at each time step by using any currently available snapshots of a dynamic network, is now attracting much attention. Nevertheless, the existing methods still face two main challenges.

\textbf{The first main challenge is the time and space complexity}. It has been discussed a lot in the static network embedding methods \cite{perozzi2014deepwalk,tang2015line,grover2016node2vec} for handling a large-scale static network. And it becomes more severe for a dynamic network. For example, assuming some small changes occur at each time step, after evolving over $k$ steps, the nodes and links over $k$ snapshots would be about $k$ times the number of them in the initial snapshot. Besides, if a real-world application requires to run downstream tasks e.g. daily, dynamic network embedding methods will therefore have the real-time constraint.

The earlier work \cite{zhu2016scalable} proposes BCGD-global that uses all available snapshots and BCGD-local that uses current snapshot and one previous snapshot. Although BCGD-local reduces the space complexity, both of them are still time-consuming, since they iteratively optimize embeddings for all nodes over each employed snapshot until convergence.
DynGEM \cite{goyal2017dyngem} proposes a heuristic strategy to modify the structure of a deep auto-encoder based on the size of the current snapshot. It continuously trains the adaptive auto-encoder using all links in current snapshot, which is also time-consuming.
NetWalk \cite{yu2018netwalk} adopts random walks to produce network walks for training the auto-encoder that minimizes the pairwise distance among all nodes in each walk. Unfortunately, it needs to retrain almost all network walks if there are a few links removed.
DynTriad \cite{zhou2018dynamic} models the triadic closure process, social homophily, and temporal smoothness in its objective function. However, optimizing the objective function requires all the links over all available snapshots.
In summary, these methods are not scalable to a large-scale dynamic network, and are intractable to fulfill real-time constraint if needed. 

\begin{figure}[htbp]
    \centering
    \includegraphics[width=0.9\textwidth]{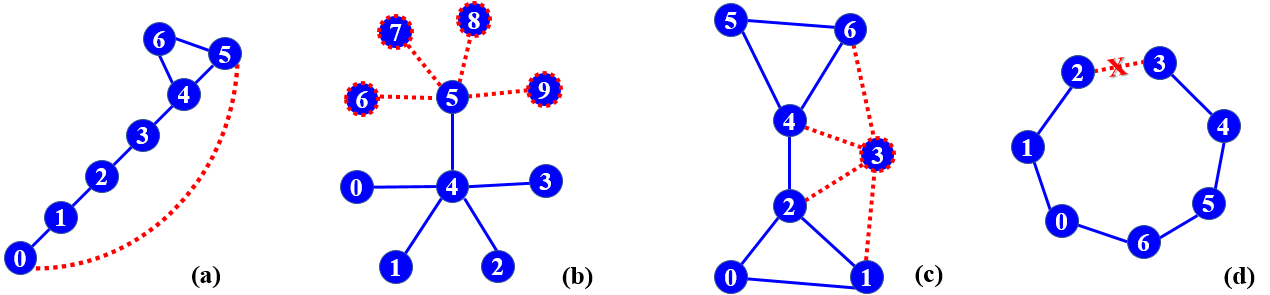}
    \caption{The changes (in red) can affect global topology of a dynamic network: (a) shows a new link is added between node 0 and 5, which also affects the proximity between node 1 and 6; (b) shows the four unseen nodes connect to node 5, which alters the global topology e.g. the structural role of node 5 is previously similar to node 0-3, but is now similar to node 4; (c) shows an unseen node 3 connects to the four seen nodes, which leads to more cliques, whereas the previous snapshot has two cliques; (d) shows an existing link is removed, which alters the global topology from a circle to a line.}
    \label{Fig1}
\end{figure}

To overcome the first main challenge, one natural idea is to put the limited computational resources on some most important parts. Recently, some methods \cite{du2018dynamic,liu2018streaming,mahdavi2018dynnode2vec} propose different strategies to select the $k$ most affected nodes, and then update embeddings for these nodes at each time step. The $k$ most affected nodes highly depend on incoming changes, and any changes can affect all nodes in a connected network via high-order proximity. Hence, the global topology of the network will be significantly altered by accumulated changes or small special changes as illustrated in Figure \ref{Fig1}, which is \textbf{the second main challenge i.e. the ability of global topology preserving}. Moreover, DHEP \cite{zhu2018high} extends HOPE \cite{ou2016asymmetric} by modifying the $k$ most affected eigenvectors based on matrix perturbation theory, so as to efficiently cope with dynamic changes. But there will be non-ignorable accumulated errors after multiple time steps \cite{zhang2018timers}, which might also be related to the second challenge.

To tackle the above two main challenges, we propose a dynamic network embedding method called DynWalks, which has the following desired properties: 1) the excellent time and space efficiency, 2) the flexibility to balance global topology and recent changes, 3) fulfilling the real-time constraint if needed, and 4) handling unseen nodes without placeholders or knowing them in advance.

The main contributions of this work are three-folds: 1) Two important challenges of dynamic network embedding are identified especially the global topology one; 2) DynWalks has some desired properties as listed above. In particular, a novel online node selecting scheme is proposal to balance global topology and recent changes; and 3) The more realistic slicing approach based on the calendar day(s) is used to slice each dynamic network. The empirical studies on six real-world dynamic networks under three different slicing ways confirm the effectiveness and efficiency of DynWalks.

\section{Notation and Problem Definition}
%In this section, a formal definition of dynamic network embedding problem will be presented.
\textit{Definition 1. Static Network Embedding:} Let $G=(\mathcal{V},\mathcal{E})$ be a given network/graph, where $\mathcal{V}$ denotes a set of $|\mathcal{V}|$ nodes/vertices and $\mathcal{E}$ denotes a set of $|\mathcal{E}|$ links/edges. 
The static network embedding aims to find a mapping function $Z=f(G)$ where $Z\in\mathbb{R}^{|\mathcal{V}|\times d}$, $d \ll |\mathcal{V}|$ and each row vector $Z_i\in\mathbb{R}^d$ is the node embedding vector corresponding to node $v_i$. The pairwise similarity of node embeddings in $Z$ should reflect the pairwise topological similarity of the nodes in $G$. 

\textit{Definition 2. The snapshots of a dynamic network:} A dynamic network in this paper refers to a time-evolving network i.e. nodes or edges might be added or removed as time goes on. The snapshots of a dynamic network are the copy of the network at each time step. A dynamic network $\mathcal{G}$ contains a set of snapshots $\mathcal{G}=(G^0,G^1,...,G^t,G^{t+1},...)$ where $t$ denotes the current time step.

\textit{Definition 3. Dynamic Network Embedding:} The dynamic network embedding problem can be generally defined as using any currently available snapshots to obtain embedding vector $Z^t_i$ for each node $v_i$ in the current snapshot $G^t$ i.e. $Z^t=f(G^t,G^{t-1},...,G^{0})$.
% Note some methods BCGD-global, DynTriad assume the future is also known so as to capture the dynamic changes??? .... but which is not realistic in real-world...   Also note that, they said they can capture the temporal trend, but we think there will be concept drift ... e.g. so it is not wise to use to long time temporal tread...
% Note that, a static embedding method like DeepWalk can be used in dynamic network embedding problem by treating each snapshot as a static network i.e. $Z^t=f(G^t)$. However, it is time-consuming.

\section{The Proposed Method}
The proposed method DynWalks includes two key components. Firstly, an online network embedding framework aims to efficiently learn embeddings based on the selected nodes. Secondly, an online node selecting scheme selects the nodes by considering both recent changes and global topology.

\subsection{Online Network Embedding Framework}
\begin{figure}[htbp]
    \centering
    \includegraphics[width=0.88\textwidth]{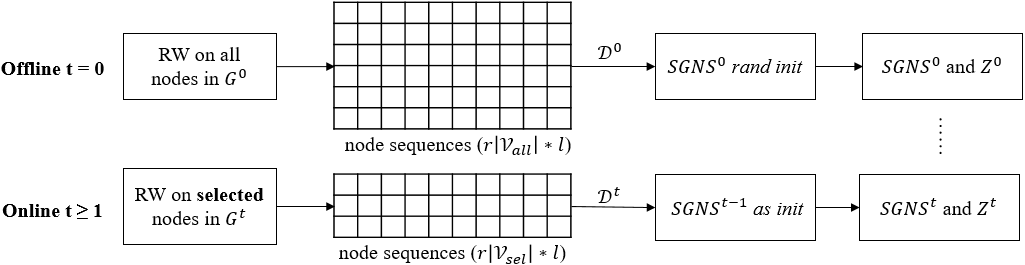}
    \caption{The overview of the proposed online network embedding framework}
    \label{Fig2}
\end{figure}
In general, the idea of capturing network topology and learning node embeddings is motivated by DeepWalk \cite{perozzi2014deepwalk}. The truncated random walks with length $l$ are conducted on each interested nodes for $r$ times, which generates $r|\mathcal{V}|$ node sequences. A sliding window with length $w+1+w$ is used to slide along each sequence, and the training pairs set $\mathcal{D}$ is built via $(v_{center}, v_{center+i})$ where $i \in [-w,+w], i \neq 0$. Note that, the random walks with length $l$ can capture at most $l$-hops away neighbors of a given starting node; and the resulting training pairs can encode $1^{st}\sim w^{th}$-order proximity of a given center node. Due to them, DynWalks can capture the high-order proximity.

In order to further reduce complexity \cite{grover2016node2vec} and freely deal with unseen nodes without placeholders, the Skip-Gram Negative Sampling (SGNS) neural network model (which is different from the model used in DeepWalk) is employed to train node embeddings over each node pair $(v_i,v_j) \in \mathcal{D}$ i.e.
\begin{equation}
    \max~~log\sigma (Z_i\cdot Z_j) + m\cdot \mathbf{E}_{v_k\sim P_{\mathcal{D}}}[log\sigma (-Z_i\cdot Z_k)]
\label{eq1}
\end{equation}
where $\sigma$ is the Sigmoid function, $Z_i$ is the node embedding vector for node $v_i$, $m$ is the number of negative samples, and $v_k$ is the negative sample from the unigram distribution $P_{\mathcal{D}}$ \cite{levy2014neural}. The aim of Eq. (\ref{eq1}) is to make the node embedding vectors similar if the nodes co-occur, and dissimilar if they are negative samples. The overall objective is to sum over all pairs i.e.
$\sum_{(v_i,v_j) \in \mathcal{D}} \#(v_i,v_j) Eq.(\ref{eq1})$. Intuitively, the more frequently a pair of nodes co-occurs, the more similar they are.

With the above modified DeepWalk i.e. a scalable static network embedding method, DynWalks further extends it to \textbf{an online manner} as shown in Figure \ref{Fig2}. It can be formalized as:
\begin{equation}
    Z^t=\left\{
    \begin{array}{lcl}
    f(G^t,SGNS^{t}_{rand}) &{t=0}\\
    f(G^t,G^{t-1},SGNS^{t-1})=f(\Delta \mathcal{E}^t,SGNS^{t-1}) &{t\geq1}
    \end{array} \right.
    \label{eq2}
\end{equation}
where $SGNS^{t-1}$ is the trained model from last time step, $Z^{t}$ is the current embedding matrix directly taken from newly trained $SGNS^{t}$ via an index operator, and streaming links $\Delta \mathcal{E}^t$ are the differences between $G^{t}$ and $G^{t-1}$ i.e. a set of triples indicating node pairs being added or removed.

As shown in Figure \ref{Fig2} and Eq. (\ref{eq2}), DynWalks has two stages. For $t=0$, random walks are conducted on all nodes in $G^0$ and $SGNS^0$ is randomly initialized, since there is no historical information of all nodes and their embeddings. For $t\geq1$, random walks are conducted on the \textbf{selected nodes} from $G^t$ where the number of nodes $|\mathcal{V}^t_{sel}|<|\mathcal{V}^t_{all}|$, so as to reduce the time and space complexity. And $SGNS^{t-1}$ is used as the initialization of $SGNS^t$, so as to utilize the historical/temporal information and stabilize the embedding vector of a node appeared in both $G^{t-1}$ and $G^{t}$.

\subsection{Online Node Selecting Scheme}
\begin{figure}[htbp]
    \centering
    \includegraphics[width=0.95\textwidth]{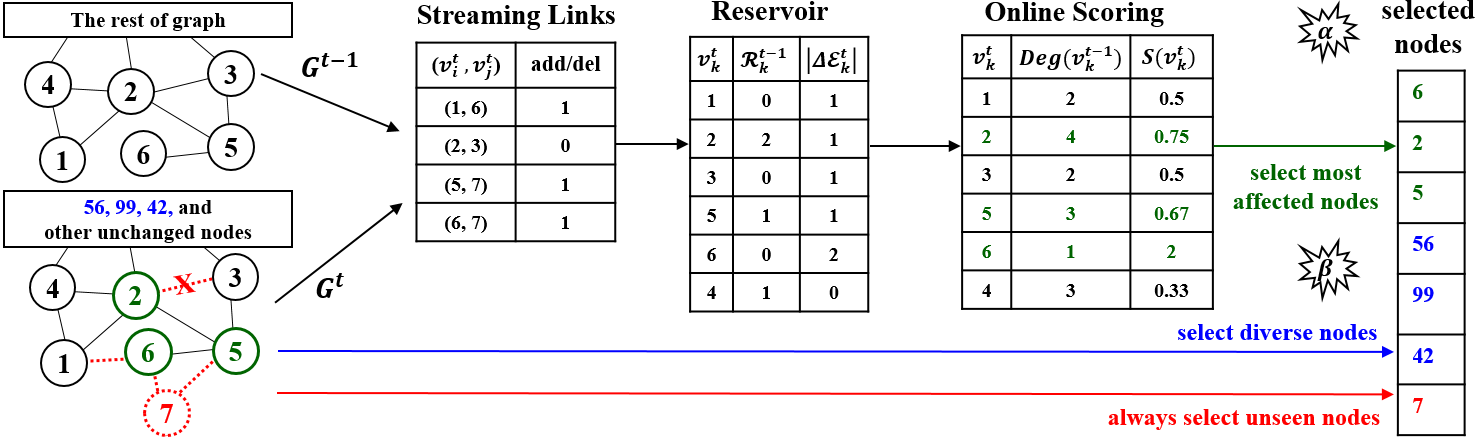}
    \caption{The illustration of the proposed online node selecting scheme}
    \label{Fig3}
\end{figure}
The online node selecting scheme as shown in Figure \ref{Fig3} is proposed to select nodes from current snapshot $G^t$. There are two questions to be answered. Firstly, how many nodes should be selected? To reduce time and space complexity, one natural idea is to focus on a part of important nodes. Therefore, the hyper-parameter $\alpha$ is used to limit the number of selected nodes to $\alpha|\mathcal{V}^t|$ where $\alpha$ can be adapted according to the real-time constraint if needed. And secondly, which nodes should be selected? To balance recent changes and global topology, the hyper-parameter $\beta$ is used to select $\beta\alpha|\mathcal{V}^t_{all}|$ \textbf{most affected nodes} and $(1-\beta)\alpha|\mathcal{V}^t_{all}|$ \textbf{diverse nodes}.

For the most affected nodes, the stream links $\Delta \mathcal{E}^t$ can be obtained based on the differences between $G^{t-1}$ and $G^t$ if not given. After that, the reservoir $\mathcal{R}$ is used to count and maintain the accumulated changes for the nodes with at least one change, e.g. $(v^t_i,v^t_j,add/del)$ gives the accumulated changes of both nodes increased by one (note the selected nodes will be removed from $\mathcal{R}$ to save space usage). Finally, motivated by the concept of inertia from Physics, the online scoring function is defined as
%\begin{footnotesize}
\begin{equation}
    S(v^t_k)
     = \frac{|\Delta\mathcal{E}^{t}_{k}|+\mathcal{R}^{t-1}_{k}}{Deg(v^{t-1}_k)} 
     = \frac{{|~\mathcal{N}({v_k^t}) \cup \mathcal{N}({v_k^{t - 1}}) - \mathcal{N}({v_k^t}) \cap \mathcal{N}({v_k^{t - 1}})}~| 
     + \mathcal{R}_k^{t-1}}{Deg(v^{t-1}_k)}
\label{eq3}
\end{equation}
%\end{footnotesize}
where $\mathcal{R}^{t-1}_{k}$ denotes the accumulated changes\footnote{The accumulated changes in reservoir are used to fix the case when a node has small changes at each time step for a long time, which will affect network topology but may be ignored if the small changes are not recorded.} of node $v_i$ at time step $t-1$, $\mathcal{N}(\cdot)$ denotes a set of the neighbors of a node, and $|\cdot|$ denotes the number of elements in a set. According to the scores of the affected nodes, the top-$k$ most affected nodes will be selected, and the time-variant $k=\beta\alpha|\mathcal{V}^t|$. Intuitively, the larger degree of a node is, the larger accumulated changes are required to select it.

For the diverse nodes, the most affected nodes in a set $\mathcal{V}_{m}^t$ as well as unseen nodes in a set $\mathcal{V}_{u}^t$ will be stored in a set of tabu nodes $\mathcal{V}_{tabu}^t=\mathcal{V}_{m}^t \cup \mathcal{V}_{u}^t$. The diverse nodes, $(1-\beta)\alpha|\mathcal{V}^t|$ in total, are then selected randomly from the set given by $\mathcal{V}_{d}^t=\mathcal{V}_{all}^t - \mathcal{V}_{tabu}^t$. Note that, other more advanced diversity-based heuristic selecting approach can be applied, which leaves as the future work.

\subsection{Algorithm and Complexity Analysis}

\setlength{\textfloatsep}{10pt}% Remove \textfloatsep
\begin{algorithm}[htbp]
\caption{DynWalks}
\label{alg1}
\textbf{Input}: network snapshots $G^0...G^{t-1},G^t$...; parameter $\alpha$ for upper limit of selected nodes; parameter $\beta$ for global topology awareness; walks per node $r$; walk length $l$; window size $w$; dimensionality $d$ \\ 
\textbf{Output}: embedding matrix $Z^t \in \mathbb{R}^{|\mathcal{V}^t| \times d}$ at each time step
\begin{algorithmic}[1] %[1] enables line numbers
\For{$t=0$} 
\State conduct random walks with length $l$ on each node in $\mathcal{V}^0_{all}$ for $r$ times
\State build node-pairs training set $\mathcal{D}^0$ based on each sliding window with size $w$ along each walk
\State initialize model $SGNS^0_{rand}$ and train it using $\mathcal{D}^0$, which returns $SGNS^0$ and $Z^0$
\EndFor
\For{$t\geq1$}
\State read streaming links set $\Delta \mathcal{E}^t$ (or obtain it by differences between $G^{t-1}$ and $G^{t}$ if not given)
\State update reservoir dictionary via $\mathcal{R}^t_{v_i}=|\Delta\mathcal{E}^{t}_{v_i}|+\mathcal{R}^{t-1}_{v_i}$ for accumulating new changes of $v_i$
\State compute score of each node in $\mathcal{R}^t$ according to Eq. (\ref{eq3}), and identify \textbf{unseen nodes} $\mathcal{V}_u^t$
\State select $(1-\beta)\alpha|\mathcal{V}_{all}^t|$ \textbf{most affected nodes} $\mathcal{V}_m^t$ based on the scores
\State store tabu nodes in set $\mathcal{V}_{tabu}^t=\mathcal{V}^t_m \cup \mathcal{V}^t_u$
\State select $\beta\alpha|\mathcal{V}_{all}^t|$ \textbf{diverse nodes} $\mathcal{V}_d^t$ by randomly sampling from the set $(\mathcal{V}^t_{all}-\mathcal{V}_{tabu}^t)$
\State remove selected nodes $\mathcal{V}^t_{sel}=\mathcal{V}^t_{u}+\mathcal{V}^t_{m}+\mathcal{V}^t_{d}$ from reservoir $\mathcal{R}^t$
\State conduct random walks with length $l$ on each node in $\mathcal{V}^t_{sel}$ for $r$ times
\State build node-pairs training set $\mathcal{D}^t$ based on each sliding window with size $w$ along each walk
\State initialize $SGNS^{t}_{init}=SGNS^{t-1}$ and train it using $\mathcal{D}^t$, which returns $SGNS^t$ and $Z^t$
\EndFor
\end{algorithmic}
\end{algorithm} %需要一些小技巧去避免极端情况，比如min(xx)，比如如何保证所有nodes个数一致。。。在appendix里面写
The pseudocode of DynWalks is summarized in Algorithm \ref{alg1}, and the source code is also provided in the supplementary materials. For $t=0$ i.e. offline stage, it is a modified version of DeepWalk \cite{perozzi2014deepwalk} which instead trains a SGNS model, such that the time complexity is further reduced to $O(r|\mathcal{V}^0_{all}|lw)$. 
For $t\geq0$ i.e. online stage, lines 13-15 are similar to lines 2-4, and the time complexity is similarly as $O(r|\mathcal{V}^t_{sel}|lw)$. In lines 6-12 i.e. the online node selecting scheme, lines 6-7 require $O(|\Delta\mathcal{E}^t|)$ respectively, and lines 8-12 at most require $O(|\mathcal{V}_{all}^t|)$ respectively. Note that, in line 9, the \textit{introselect} algorithm is used to find the element in the top-$k$ position without fully sorting other elements and then use it to select the top-$k$ values, which gives $O(|\mathcal{V}_{all}^t|)$. Therefore, the overall time complexity during online stage is $O(2|\Delta\mathcal{E}^t|+5|\mathcal{V}_{all}^t|+r|\mathcal{V}^t_{sel}|lw)$. 
For most real-world large dynamic networks, the streaming links $|\Delta\mathcal{E}^t|\approx c|\mathcal{V}_{all}^t|$ where $c$ is usually a constant smaller than the average degree, and the selected nodes $|\mathcal{V}^t_{sel}|\approx \alpha |\mathcal{V}^t_{all}|$ due to relatively less unseen nodes. The overall time complexity during online stage can be rewritten as $O((2c+5+\alpha rlw) \cdot |\mathcal{V}^t_{all}|)$ where $\alpha \in [0.0, 1.0]$. To meet the real-time constraint if needed, one may decrease $\alpha$ at the risk of decreasing performance. %but the experiments in this paper show that $\alpha=0.2$ can already obtain satisfactory results. 

Regarding the space complexity, there are four space-consuming variables kept in memory: the previous SGNS model with size $2|\mathcal{V}^{t-1}_{all}|d$, the reservoir dictionary with size at most $|\mathcal{V}^t_{all}|$, the streaming edges $\Delta\mathcal{E}^t$ or two snapshots $G^{t-1}$ and $G^t$, and the training pairs $|\mathcal{D}^t|= O(r|\mathcal{V}^t_{sel}|wl)$. Therefore, the space complexity is linearly proportional to the number of nodes in $G$.

\subsection{Global Topology and Recent Changes Awareness}
\begin{figure}[htbp]
    \centering
    \includegraphics[width=0.9999\textwidth]{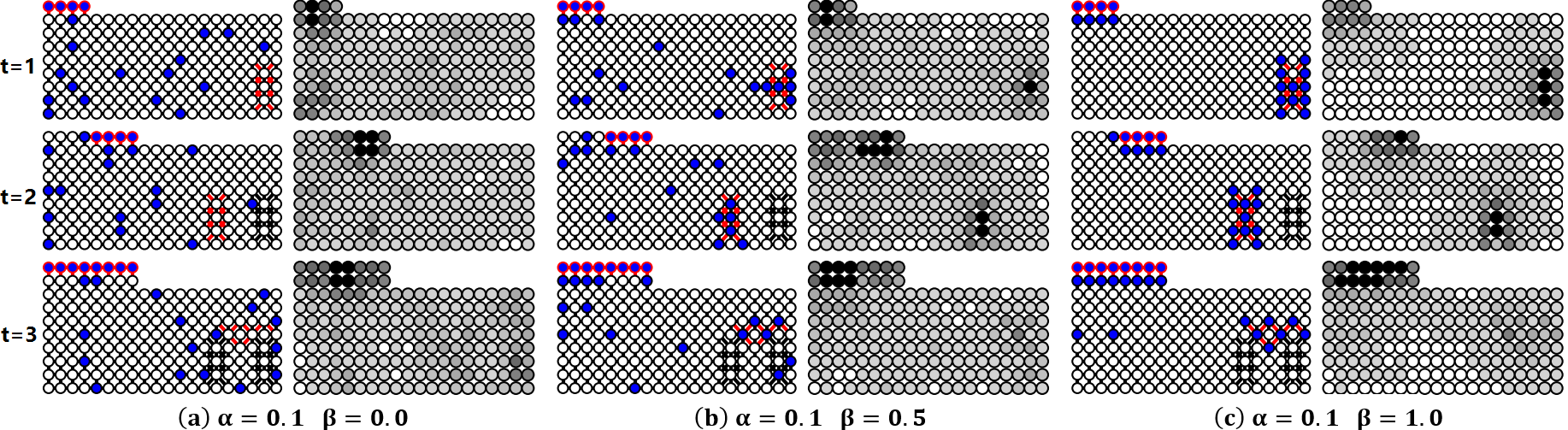}
    \caption{Balancing global topology and recent changes: the left ones of each sub-figure show the selected nodes in blue, and new nodes/links are in red; the right ones show the number of times that nodes are trained in grey-scale. At $t=0$, the toy network is a regular 2d-grid graph with size (20, 8).}
    \label{Fig4}
\end{figure} %可以考虑把自己图里面间距缩小！！！然后下方让出来
DynWalks can aware and balance global topology and recent changes via the hyper-parameter $\beta$. As shown in Figure \ref{Fig4} (except the red unseen nodes at the top-left corner obviously needed to select and train), for $\beta=0.0$ in (a) i.e. focus on global topology, the selected nodes in blue are diversely distributed; whereas for $\beta=1.0$ in (c) i.e. focus on recent changes, all the recently changed nodes by new links in red (roughly at the middle-right corner) are selected. Consequently, the grey-scale heatmap of (a) implies almost all nodes get similar attention for training; whereas the heatmap of (c) shows the training is bias to recent changes, but neglects other parts i.e. the large light parts in the heatmaps. Nevertheless, choosing a proper $\beta$ in between e.g. $\beta=0.5$ in (b), a good trade-off between global topology and recent changes may be achieved. Note that the darker a node is, the more frequently it occurs in $\mathcal{D}$, and hence it is trained by Eq. (\ref{eq1}) for more times.

% Proof the almost surely of global network updating..
% It is easy to prove that after about 1/(a*b) time steps, all nodes of a network will be update for one round with probability one.... which gurantee the glotbal toplogy.. by at least update each node....

\section{Empirical Study}
The six real-world dynamic networks come from \url{http://konect.uni-koblenz.de/} or \url{http://snap.stanford.edu/data/}. DNC is the leaked email communication network of US Democratic National Committee in 2016. AS733 contains 733 daily instances of the Autonomous System of routers exchanging traffic flows with neighbors. Chess is an online gaming network of Chess where each pair of nodes records a game between two players. Elec is the network of English Wikipedia users vote for and against each other in admin elections. FBW is a social network of Facebook Wall posts where nodes are the users and links are built based on the interactions in their wall posts. HepPh is a co-author network extracted from the papers of High Energy Phsics Phenomenology in the arXiv. All datasets except AS733 are given in the streaming links format i.e. a pair of nodes with a timestamp (streaming links may bring unseen nodes), whereas AS733 is given in the snapshots format and only AS733 has nodes/links being removed. All datasets are treated as undirected networks. Besides, the minimum possible interval in Chess is 1 month, whereas the interval for other datasets is 1 day. Moreover, three ways are employed to slice dynamic networks into snapshots as shown in Table \ref{Tab1}, and only the most recent snapshots if appropriate are taken out after data slicing (the source code is provided in supplementary materials): \textbf{S1}--the minimum interval of each dataset for 21 snapshots, \textbf{S2}--the maximum possible interval for 21 snapshots, \textbf{S3}--the minimum interval for 100 snapshots.
% n/a 要么如AS733分割30天没太大意义；要么如S3，其他数据跑不了。。
% 需要去掉S3-DNC吗？没有跑。。。

\begin{table}[htbp]
  \centering
  \renewcommand\tabcolsep{5.3pt}
  \caption{The statistics of six real-world dynamic networks under three slicing ways}
    \scalebox{0.92}{
    \begin{tabular}{ll|cc|cc|cc}
    \toprule
          &   \textbf{S1, S2, S3}    & \multicolumn{2}{c|}{21 SnapShots} & \multicolumn{2}{c|}{21 SnapShots} & \multicolumn{2}{c}{100 SnapShots} \\
    Data  & Intervals & Nodes (k) & Links (k) & Nodes (k) & Links (k) & Nodes (k) & Links (k) \\
    \midrule
    DNC   & 1d, 10d, 1d & \textbf{1}.02-\textbf{1}.85 & \textbf{2}.273-\textbf{4}.330 & \textbf{0}.06-\textbf{1}.18 & \textbf{0}.062-\textbf{2}.752 & n/a & n/a \\
    AS733  & 1d, n/a, 1d & \textbf{1}.48-\textbf{3}.57 & \textbf{3}.132-\textbf{7}.033 & n/a   & n/a   & \textbf{0}.10-\textbf{6}.47 & \textbf{0}.239-\textbf{12}.57 \\
    Chess  & 1m, 4m, 1m & \textbf{4}.39-\textbf{7}.05 & \textbf{33}.11-\textbf{52}.34 & \textbf{1}.92-\textbf{5}.15 & \textbf{6}.404-\textbf{40}.62 & \textbf{0}.29-\textbf{7}.30 & \textbf{0}.564-\textbf{55}.90 \\
    Elec  & 1d, 30d, 1d & \textbf{7}.02-\textbf{7}.10 & \textbf{98}.97-\textbf{100}.5 & \textbf{2}.89-\textbf{6}.83 & \textbf{31}.82-\textbf{96}.21 & n/a   & n/a \\
    FBW   & 1d, 30d, n/a & \textbf{43}.7-\textbf{45}.8 & \textbf{170}.6-\textbf{183}.0 & \textbf{11}.5-\textbf{41}.8 & \textbf{35}.48-\textbf{162}.7 & n/a   & n/a \\
    HepPh  & 1d, 30d, n/a & \textbf{16}.8-\textbf{17}.0 & \textbf{1171}.-\textbf{1194}. & \textbf{10}.7-\textbf{16}.4 & \textbf{575}.1-\textbf{1126}. & n/a   & n/a \\
    \bottomrule
    \end{tabular}%
    }
  \label{Tab1}%
\end{table}

In the experiments, the proposed method is compared to four state-of-the-art dynamic network embedding methods. BCGD-local \cite{zhu2016scalable}, DynGEM \cite{goyal2017dyngem}, and DynWalks (Ours) are online learning method, as they only use the topological information available at current time step to learning embeddings for current time step, whereas BCGD-global \cite{zhu2016scalable}  and DynTriad \cite{zhou2018dynamic} use all topological information including future time steps together to learn embeddings for each time step. The original source code in \url{https://github.com/linhongseba/Temporal-Network-Embedding}, \url{https://github.com/palash1992/DynamicGEM}, and \url{https://github.com/luckiezhou/DynamicTriad} for BCGD, DynGEM, and DynTriad respectively is used. And the original hyper-parameters in the source code are employed and fixed for all experiments. For fairness, the key hyper-parameters of our method are also fixed at $\alpha=0.2$ and $\beta=0.5$ for all experiments, and other parameters $r$, $l$, $w$ are fixed at 20, 80, 10 respectively \cite{perozzi2014deepwalk}. The embedding size $d$ is set to 128 for all methods.

\subsection{Graph Reconstruction and Link Prediction}
Graph Reconstruction (GR) is used to evaluate the quality of the embeddings for recovering the network topology of the interested nodes. In this paper, GR task aims to retrieve for $1/4$ randomly sampled nodes of the snapshot at $t$ using embeddings at $t$, whereas Changed GR (CGR) task aims to retrieve for the nodes directly affected by changes. Both tasks are then evaluated by Average Precision@k (AP@k) score \cite{wang2016structural}. Link Prediction (LP) is used to evaluate the ability of predicting future links at $t+1$ using embeddings at $t$. The testing links include all changed links from $t$ to $t+1$, plus other links randomly sampled from the snapshot at $t+1$ for balancing existent links and non-existent links. LP task is then evaluated by Area under the ROC Curve (AUC) score \cite{zhu2016scalable}.

\begin{table}[htbp]
  \centering
  \setlength{\abovecaptionskip}{0pt}
  \setlength{\belowcaptionskip}{6pt}
  \renewcommand\tabcolsep{4.4pt}
  \caption{CGR, GR and LP tasks under slicing ways S1 (left) and S2 (right): each entry (in $\%$) is calculated by the mean over 20 time steps and over 10 runs; the best result of each half row is in bold.}
    \scalebox{0.92}{
    \begin{tabular}{l|ccccc|ccccc}
    \toprule
          & GEM   & BCGD$^{l}$ & BCGD$^{g}$ & Triad & Ours$^{0.2,0.5}$ & GEM   & BCGD$^{l}$ & BCGD$^{g}$ & Triad & Ours$^{0.2,0.5}$ \\
    \midrule
          & \multicolumn{5}{c|}{CGR-AP@10-S1}     & \multicolumn{5}{c}{CGR-AP@10-S2} \\
    \midrule
    AS733 & 06.38  & 53.90  & 13.87 & 56.39 & \textbf{79.19} & n/a   & n/a   & n/a   & n/a   & n/a \\
    Chess & 12.10  & 46.99 & 15.85 & 50.27 & \textbf{80.27} & 08.56  & 44.81 & 11.68 & 52.57 & \textbf{81.40} \\
    DNC   & 25.55 & 57.50  & 49.02 & \textbf{67.42} & 65.85 & n/a   & 38.03 & 73.67 & \textbf{83.27} & 81.88 \\
    Elec  & n/a   & 38.57 & 29.89 & 46.07 & \textbf{65.11} & 12.19 & 26.45 & 19.94 & 49.39 & \textbf{58.57} \\
    FBW    & n/a   & 06.66  & 00.22  & 51.60  & \textbf{85.08} & n/a   & 07.16  & 00.22  & 62.97 & \textbf{86.79} \\
    HepPh & n/a   & 74.31 & 56.62 & n/a   & \textbf{84.81} & n/a   & 67.20  & 46.90  & n/a   & \textbf{83.50} \\
    \midrule
          & \multicolumn{5}{c|}{GR-AP@10-S1}      & \multicolumn{5}{c}{GR-AP@10-S2} \\
    \midrule
    AS733 & 00.60   & 48.83 & 02.48  & 63.31 & \textbf{81.12} & n/a   & n/a   & n/a   & n/a   & n/a \\
    Chess & 04.41  & 43.84 & 04.41  & 54.61 & \textbf{85.80} & 04.94  & 49.74 & 04.74  & 57.72 & \textbf{87.05} \\
    DNC   & 03.33  & 34.96 & 22.14 & \textbf{76.93} & 76.02 & n/a   & 12.09 & 75.92 & \textbf{81.72} & 51.26 \\
    Elec  & n/a   & 17.89 & 09.13  & 57.71 & \textbf{81.62} & 03.85  & 16.31 & 08.94  & 59.65 & \textbf{78.27} \\
    FBW    & n/a   & 03.94  & 00.11  & 58.94 & \textbf{90.20} & n/a   & 04.02  & 00.14  & 71.29 & \textbf{91.42} \\
    HepPh & n/a   & 61.34 & 31.40  & n/a   & \textbf{81.27} & n/a   & 55.55 & 27.72 & n/a   & \textbf{80.61} \\
    \midrule
          & \multicolumn{5}{c|}{CGR-AP@100-S1}    & \multicolumn{5}{c}{CGR-AP@100-S2} \\
    \midrule
    AS733 & 06.67  & 80.42 & 80.35 & 73.34 & \textbf{91.30} & n/a   & n/a   & n/a   & n/a   & n/a \\
    Chess & 13.06 & 64.26 & 40.82 & 61.12 & \textbf{84.68} & 11.89 & 71.83 & 51.93 & 67.12 & \textbf{90.66} \\
    DNC   & 29.24 & 82.25 & 75.28 & 83.48 & \textbf{87.14} & n/a   & 73.01 & 79.92 & 95.18 & \textbf{99.31} \\
    Elec  & n/a   & 48.45 & 38.93 & 56.04 & \textbf{69.40} & 12.47 & 39.33 & 33.54 & 61.68 & \textbf{66.88} \\
    FBW    & n/a   & 06.91  & 00.23  & 63.00    & \textbf{97.00} & n/a   & 11.45 & 02.50   & 76.47 & \textbf{98.68} \\
    HepPh & n/a   & 68.70  & 53.88 & n/a   & \textbf{85.51} & n/a   & 62.21 & 45.87 & n/a   & \textbf{84.51} \\
    \midrule
          & \multicolumn{5}{c|}{GR-AP@100-S1}     & \multicolumn{5}{c}{ GR-AP@100-S2} \\
    \midrule
    AS733 & 01.31  & 88.87 & 95.39 & 83.15 & \textbf{97.15} & n/a   & n/a   & n/a   & n/a   & n/a \\
    Chess & 06.55  & 74.81 & 45.51 & 71.83 & \textbf{94.90} & 08.91  & 82.25 & 59.78 & 74.99 & \textbf{96.79} \\
    DNC   & 05.77  & 81.45 & 82.84 & 94.38 & \textbf{97.11} & n/a   & 82.51 & 88.54 & 98.42 & \textbf{99.85} \\
    Elec  & n/a   & 42.29 & 36.75 & 74.07 & \textbf{86.74} & 04.55  & 40.10  & 41.57 & 76.17 & \textbf{86.05} \\
    FBW    & n/a   & 04.88  & 00.17  & 71.42 & \textbf{98.59} & n/a   & 08.60   & 02.69  & 83.52 & \textbf{99.39} \\
    HepPh & n/a   & 58.28 & 30.17 & n/a   & \textbf{84.91} & n/a   & 55.84 & 30.96 & n/a   & \textbf{85.83} \\
    \midrule
          & \multicolumn{5}{c|}{LP-AUC-S1}        & \multicolumn{5}{c}{ LP-AUC-S2} \\
    \midrule
    AS733 & 60.18 & 61.37 & 70.15 & 65.54 & \textbf{85.56} & n/a   & n/a   & n/a   & n/a   & n/a \\
    Chess & 64.23 & \textbf{86.66} & 85.77 & 79.32 & 77.24 & 68.79 & \textbf{88.62} & 79.94 & 85.83 & 73.12 \\
    DNC   & 75.90  & 84.18 & 89.34 & \textbf{90.30} & 78.76 & n/a   & 76.52 & \textbf{94.21} & 92.81 & 89.82 \\
    Elec  & 66.32 & 91.16 & 82.83 & \textbf{97.31} & 90.31 & 67.69 & 82.13 & 82.47 & \textbf{90.34} & 78.38 \\
    FBW    & n/a   & 82.83 & 82.88 & 81.76 & \textbf{88.00} & n/a   & 84.51 & 85.02 & 83.93 & \textbf{89.69} \\
    HepPh & n/a   & 88.39 & 82.37 & n/a   & \textbf{90.25} & n/a   & \textbf{89.99} & 81.17 & n/a   & 89.58 \\
    \bottomrule
    \end{tabular}%
    }
  \label{Tab2}%
\end{table}%

% 几个任务不同的偏好，更偏向global: GR-AP@100, CGR-AP@100, GR-AP@10, CGR-AP@10, LP 
% The n/a values on S2-AS733 are due to the weak real-world meaning, and other n/a values are due to running over either 24 hours or 128G ROM.
As shown in Table \ref{Tab2}, for all CGR and GR tasks, our method significantly outperforms other methods in most cases. One exception is the GR-AP@10-S2-DNC case where BCGD-local and our method both obtain bad results, which is due to that S2 makes the snapshots of DNC snapshots dramatically sparse as shown in Table \ref{Tab1} (i.e. the email network at very beginning has 64 nodes and 62 links). DynTriad and BCGD-global can still obtain good results, however, they use all links over all snapshots together (hence overcome the sparsity problem but is not realistic in some scenarios) to learn embeddings for each time step, whereas BCGD-local and ours are the online learning method. For all LP tasks, all methods except DynGEM are comparable, since no method can always outperform others.

\begin{figure}[htbp]
    \centering
    \includegraphics[width=0.9999\textwidth]{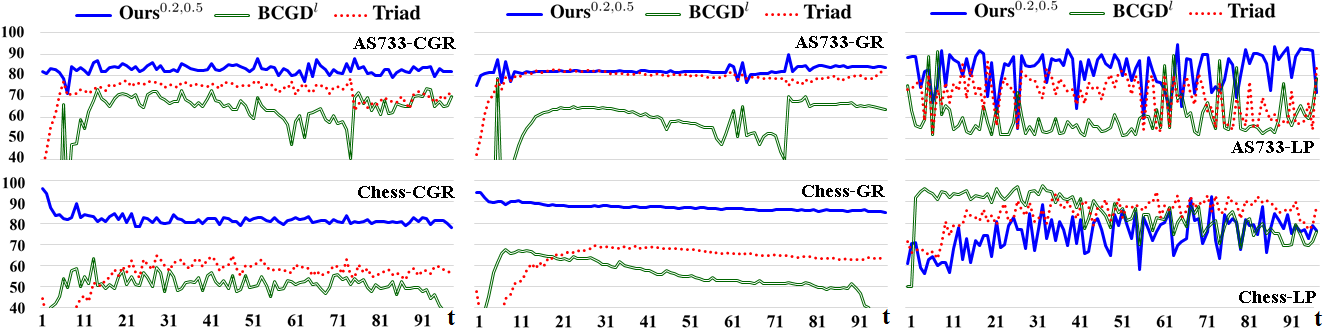}
    \caption{CGR-AP@10, GR-AP@10 and LP tasks under slicing way S3 with 100 time steps: each point for drawing the line chart is given by the mean over 10 runs at each time step along the x-axis.}
    \label{Fig5}
\end{figure}

As shown in Figure \ref{Fig5}, the methods that can complete the given tasks with 100 time steps within a reasonable duration are conducted. For the CGR and GR tasks, our method not only outperforms other methods, but also exhibits the stability w.r.t. time evolving, which may verify the usefulness of using embeddings at $t$-$1$ as the initialization of embeddings at $t$. For Chess-LP task, our method performs not that good at the beginning, but gradually catches up with others as the network grows. Furthermore, regarding all experiments, our method is better at GR/CGR tasks than LP tasks, since GR/CGR tasks ask for correctly recovering as many the neighbors of a given node as possible, which requires the good knowledge of global topology---the special advantage of our method.

% 另外long还说明了长时间不会有累积误差。。其实我们用0.2参数，几次后就可以整个图更新一遍。。。。%%%%%也说明了global topology保留的重要性。。。------------------------- 
% 在LP上我们在Chess前期干不过别人，原因是指向数据，太稀疏，cite自己文章，后面就追上了
% LP大家都比较不稳定是因为每次测试数据集少；

\subsection{Sensitivity Analysis}
\begin{figure}[htbp]
    \centering
    \includegraphics[width=0.9999\textwidth]{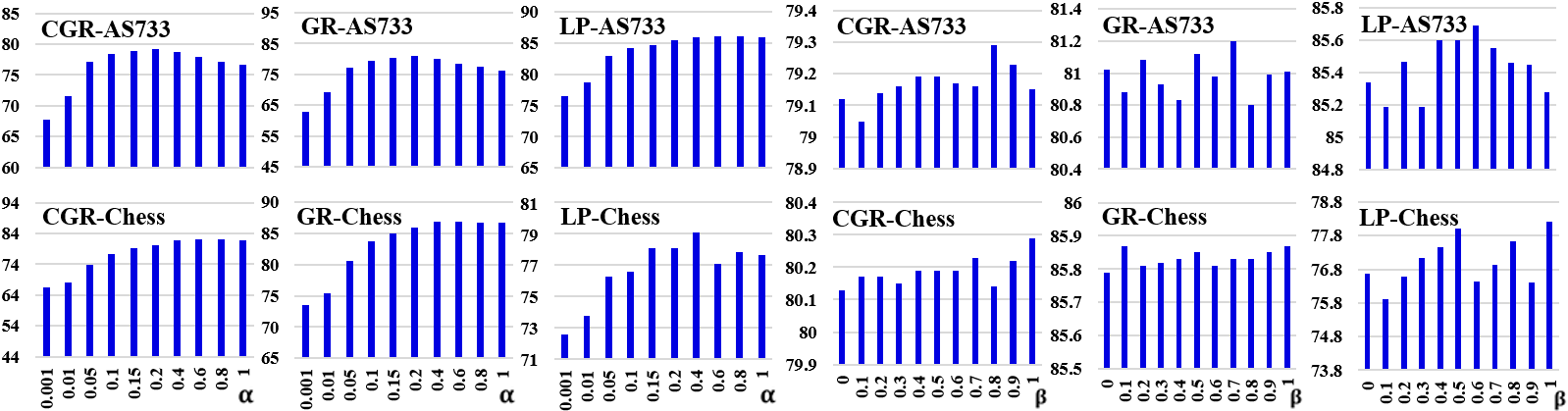}
    \caption{Sensitivity analysis over 10 runs on CGR-AP@10, GR-AP@10 and LP tasks under slicing way S1: the left six ones for $\beta=0.5$ and varying $\alpha$, and the right six ones for $\alpha=0.2$ and varying $\beta$.}
    \label{Fig6}
\end{figure}

The sensitivity analysis of two hyper-parameters $\alpha$ and $\beta$ is shown in Figure \ref{Fig6}. For $\beta=0.5$ and varying $\alpha$ i.e. to see the effect of different numbers of the selected nodes: the performance significantly increases as $\alpha$ grows when $\alpha$ is relatively small, but may saturate or decrease when $\alpha$ is relatively large. For $\alpha=0.2$ and varying $\beta$ i.e. to see the effect of different attentions between global topology and recent changes: there is no clear tendency as $\beta$ grows, since different datasets and different tasks both have different bias. Note that, $\alpha=0.2$ and $\beta=0.5$ are fixed for all other experiments, however, it is possible to obtain even better results via tuning them properly.

% 5）其实通过未调整，LP是可以上去的，比如图6，，，0.4带来了6个百分点！但是为了公平我们固定参数比较，
% 1) no need to update for all nodes;
% set a =0.2 for real-time??

\subsection{Wall-Clock Time}
To access wall-clock time, all the results as shown in Table \ref{Tab3} are produced in the same computer with 128G ROM, 40-cores Intel E5-2630v4@2.2GHz CPU, and Nvidia Tesla-P100-16G GPU. The n/a values are due to the methods exceed either memory limit or 12 hours. It is obvious that our method is much faster than other methods (DynTriad uses GPU for acceleration and hence may be slightly faster on the small dynamic network within 21 time steps). Considering online stage Ours$_{t\geq1,avg}^{0.2,0.5}$, the superiority will become more obvious if lasting for more times steps. Besides, the last three columns and Figure \ref{Fig6} show the practical significance of $\alpha$ to balance wall-clock time and performance.
%Besides, the last three columns of $\alpha=0.2, 0.1, 1.0$ verify the real-world approximation of time complexity $O(\alpha \cdot r|\mathcal{V}^t_{all}|lw)$. Together with Figure \ref{Fig6}, $\alpha$ presents the practical significance to balance performance and real-time constraint.

\begin{table}[htbp]   % Ours 1.0 0.5的时间更好！！！
  \renewcommand\tabcolsep{3.05pt}
  \centering
  \caption{Wall-clock time under slicing way S1: the left part shows the total time over 21 time steps including I/O; the right part shows the detailed time of the offline or online stage excluding I/O.}
    \scalebox{0.92}{
    \begin{tabular}{c|ccccc|cccc}
    \toprule
          & GEM & BCGD$^{l}$ & BCGD$^{g}$ & Triad & Ours$^{0.2,0.5}$ & Ours$_{t=0}^{0.2,0.5}$ & Ours$_{t\geq1,avg}^{0.2,0.5}$& Ours$_{t\geq1,avg}^{0.4,0.5}$ & Ours$_{t\geq1,avg}^{0.6,0.5}$\\
    \midrule
    DNC   & 696   & 241   & 939   & 129   & \textbf{91}    & 11.31 & 3.89 & 6.90 & 9.98 \\
    AS733 & 1289  & 563   & 1800  & \textbf{111}   & 152   & 16.31  & 6.68 & 11.40 & 16.05 \\
    Chess & 5495  & 1505  & 4108  & 776   & \textbf{333}   & 47.85 & 14.08 & 25.48 & 36.79 \\
    Elec  & 9677  & 1902  & 4829  & 2729  & \textbf{495}   & 88.57 & 20.07 & 37.80 & 55.54 \\
    HepPh & n/a   & 11878 & 16834 & n/a   & \textbf{2128}  & 319.02 & 88.71 & 152.12 & 223.54 \\
    FBW & n/a   & 10042 & 23139 & 4896  & \textbf{2506}  & 473.79 & 100.83 & 192.38 & 285.69\\
    \bottomrule
    \end{tabular}%
    }
  \label{Tab3}%
\end{table}

\section{Conclusion}
This paper first identifies the two main challenges about complexity and global topology preserving of the dynamic network embedding problem. DynWalks is hence proposed to tackle these challenges. To reduce the complexity, an online network embedding framework extends the modified DeepWalk so as to dynamically and efficiently learn embeddings based on the selected nodes. To preserve the global topology---an important feature for downstream tasks, a novel node selecting scheme offers the flexible choices to balance global topology and recent changes. The extensive empirical studies have verified the effectiveness and efficiency of DynWalks. The limitation might be the cold start problem at the very beginning of a dynamic network if the snapshot is extremely sparse, one future work is to employ side information e.g. node attributes to overcome it. Another future work is to improve the diverse nodes selecting approach \cite{bletsas2006simple} for better maintaining the global topology.  % adaptive alpha and beta???

\newpage

% Note that publication-quality tables \emph{do not contain vertical rules.} We strongly suggest the use of the \verb+booktabs+ package, which allows for typesetting high-quality, professional tables:

%\section*{Acknowledgment}
%Use unnumbered third level headings for the acknowledgments. All acknowledgments go at the end of the paper. Do not include acknowledgments in the anonymized submission, only in the final paper.
%\medskip
\small
\bibliographystyle{unsrt}
\bibliography{neurips_2019}

\normalsize
\newpage
\appendix
{\Large\textbf{Appendices}}
\section{Implementation Details of DynWalks and Other Compared Methods}
The pseudocode of DynWalks is summarized in Algorithm 1, and the source code implemented in Python is also provided in the supplementary materials. However, there is a trick not explicitly mentioned in the paper, but it can help DynWalks properly deal with either very large or few changes for any choices of $\alpha$ and $\beta$. In order to make the wall-clock time controllable (by limiting the number of selected nodes), as well as to retain an acceptable performance (by choosing as many selected nodes as possible), DynWalks asks for selecting exactly $\alpha |\mathcal{V}^t_{all}|$ nodes in total, which includes $\beta\alpha|\mathcal{V}^t_{all}|$ most affected nodes and $(1-\beta)\alpha|\mathcal{V}^t_{all}|$ diverse nodes. However, sampling $\beta\alpha|\mathcal{V}^t_{all}|$ most affected nodes from the reservoir might be impossible, if the number of nodes maintained in the reservoir is less than $\beta\alpha|\mathcal{V}^t_{all}|$. In this case, the small trick is to calculate the number of lacked nodes, which is then added to the number of diverse nodes, so that the total selected nodes are still $\alpha |\mathcal{V}^t_{all}|$. As a consequence, the performance is likely to be superior (not wasting computational resources) and the wall-clock time is also controllable (not exceeding required time). Please see the function \textit{node\_selecting\_scheme()} in \textit{source code $\rightarrow$ src $\rightarrow$ libne $\rightarrow$ DynWalks.py} for details.

In this paper, DynWalks is compared with other four dynamic network embedding methods. We re-implement these methods according to the following procedures. 1) BCGD-local, BCGD-gloabl, and DynTriad need to know all unique nodes in advance. Hence, we write a Python script to obtain all unique nodes throughout all snapshots. 2) Run the their code to obtain node embeddings, and convert node embeddings in different formats into the standard format i.e. dict-of-dict $\{t1: Z_1, t2: Z_2, ...\}$ where $Z_1=\{node1: emb1, node2: emb2, ...\}$. 3) Since all the outputs of all methods are now in the same format, we employ \textit{eval.py} and \textit{libne} as provided in supplementary materials for evaluation. 4) During running the compared methods, some methods for some tasks may exceed 128G ROM or 12 hours, which leads to the n/a values as shown in the tables. In general, DynGEM consumes very large ROM, and the complexity might be related to the number of nodes; DynTriad also consumes relatively large ROM, but the complexity might be related to the number of links; BCGD-local and BCGD-global only consume small ROM, but they learn node embeddings in an iterative way over all given snapshots until convergence, which is particularly time-consuming.

\section{Hyper-parameters Tuning of DynWalks}
There are two types of hyper-parameters: the key hyper-parameters i.e. $\alpha=0.2$ for the upper limit of selected nodes and $\beta=0.5$ for balancing global topology and recent changes; and the other hyper-parameters related to random walks and Skip-Gram model i.e. walks per node $r=20$, walk length $l=80$, window size $w=10$, embedding dimensionality $d=128$, and negative samples of SGNS model in Eq \ref{eq1} $m=5$. For fairness, we fix the above hyper-parameters for all experiments except sensitivity analysis. The experimental results in Table \ref{Tab2}, Figure \ref{Fig5}, and Table \ref{Tab3} show that the fixed hyper-parameters have already obtain the superior results in terms of both downstream tasks and walk-clock time. Nevertheless, according to the sensitivity analysis, it is possible to obtain even better results if we tune $\alpha$ and $\beta$ properly. 

The practical guidance of tuning $\alpha$ and $\beta$ according to our experience is as follows. First of all, we should choose a proper $\alpha$ to fulfill the real-time constraint based on the size of the dynamic network and the computational resources. Because the real-world approximation of time complexity is $O((2c+5+\alpha rlw) \cdot |\mathcal{V}^t_{all}|)$, we can use a toy dynamic network to obtain the wall-clock time in the machine, and then deduce the estimated wall-clock time of the large dynamic network in the same machine via $\frac{t_{large}}{t_{toy}} \approx \frac{|\mathcal{V}_{large}|}{|\mathcal{V}_{toy}|}$. 
After that, we could choose a proper $\beta$ via gird search in $[0.0, 1.0]$, since different datasets and different tasks both have different bias to global topology.

In additional, DeepWalk \cite{perozzi2014deepwalk} can be seen as a special case of DynWalks when $\alpha=1.0$ and replacing the original Skip-Gram Hierarchical Softmax model with Skip-Gram Negative Sampling model. However, according to Figure \ref{Fig6} and Table \ref{Tab3}, too large $\alpha$ brings increasing wall-clock time, but it is hard to receive further increasing performance. Based on our experience, the performance significantly increases as $\alpha$ grows when $\alpha$ is relatively small e.g. below 0.5 on AS7SS dataset.

\section{Graph Reconstruction and Link Prediction: The Standard Deviation}
\begin{table}[htbp]
  \centering
  \caption{CGR, GR and LP tasks under sliding way S1: the mean is given by the average over 20 time steps and over 10 runs; the std is given by the average of std over 10 runs; the best result is in bold.}
    \renewcommand\tabcolsep{6pt}
    \scalebox{0.92}{
    \begin{tabular}{l|cccccc}
    \toprule
          & GEM & BCGD$^{l}$ & BCGD$^{g}$ & Triad & ours$^{0.1,0.5}$ & ours$^{0.2,0.5}$ \\
    \midrule
          & \multicolumn{6}{c}{CGR-AP@10-S1 (mean $\pm$ std)} \\
    \midrule
    AS733 & 06.38$\pm$02.27 & 53.90$\pm$15.43 & 13.87$\pm$03.39 & 56.39$\pm$09.98 & 78.45$\pm$02.56 & \textbf{79.19$\pm$02.68} \\
    Chess & 12.10$\pm$04.71 & 46.99$\pm$08.06 & 15.85$\pm$07.02 & 50.27$\pm$08.39 & 77.29$\pm$02.06 & \textbf{80.27$\pm$01.39} \\
    DNC   & 25.55$\pm$06.38 & 57.50$\pm$06.24 & 49.02$\pm$07.14 & \textbf{67.42$\pm$11.63} & 64.38$\pm$05.39 & 65.85$\pm$05.44 \\
    Elec  & n/a & 38.57$\pm$03.87 & 29.89$\pm$03.41 & 46.07$\pm$04.98 & 59.51$\pm$03.83 & \textbf{65.11$\pm$03.04} \\
    FBW    & n/a & 06.66$\pm$05.37 & 00.22$\pm$00.06 & 51.60$\pm$11.60 & 83.70$\pm$00.59 & \textbf{85.08$\pm$00.58} \\
    HepPh & n/a & 74.31$\pm$11.76 & 56.62$\pm$13.37 & n/a & 82.35$\pm$04.23 & \textbf{84.81$\pm$03.54} \\
    \midrule
          & \multicolumn{6}{c}{GR-AP@10-S1 (mean $\pm$ std)} \\
    \midrule
    AS733 & 00.60$\pm$00.24 & 48.83$\pm$16.59 & 02.48$\pm$01.56 & 63.31$\pm$10.38 & 79.48$\pm$02.31 & \textbf{81.12$\pm$02.56} \\
    Chess & 04.41$\pm$00.84 & 43.84$\pm$13.43 & 04.41$\pm$00.87 & 54.61$\pm$09.64 & 83.71$\pm$00.90 & \textbf{85.80$\pm$00.46} \\
    DNC   & 03.33$\pm$00.65 & 34.96$\pm$04.28 & 22.14$\pm$10.58 & \textbf{76.93$\pm$10.91} & 71.06$\pm$05.71 & 76.02$\pm$04.93 \\
    Elec  & n/a & 17.89$\pm$03.80 & 09.13$\pm$00.45 & 57.71$\pm$07.82 & 77.80$\pm$01.69 & \textbf{81.62$\pm$01.28} \\
    FBW    & n/a & 03.94$\pm$03.41 & 00.11$\pm$00.02 & 58.94$\pm$11.75 & 89.20$\pm$00.27 & \textbf{90.20$\pm$00.14} \\
    HepPh & n/a & 61.34$\pm$08.60 & 31.40$\pm$00.57 & n/a & 77.77$\pm$01.60 & \textbf{81.27$\pm$01.03} \\
    \midrule
          & \multicolumn{6}{c}{CGR-AP@100-S1 (mean $\pm$ std)} \\
    \midrule
    AS733 & 06.67$\pm$01.78 & 80.42$\pm$07.23 & 80.35$\pm$06.44 & 73.34$\pm$06.37 & 90.45$\pm$01.09 & \textbf{91.30$\pm$01.14} \\
    Chess & 13.06$\pm$04.58 & 64.26$\pm$05.36 & 40.82$\pm$06.95 & 61.12$\pm$09.66 & 83.76$\pm$04.73 & \textbf{84.68$\pm$04.43} \\
    DNC   & 29.24$\pm$06.25 & 82.25$\pm$05.23 & 75.28$\pm$11.40 & 83.48$\pm$05.95 & 86.04$\pm$02.73 & \textbf{87.14$\pm$02.55} \\
    Elec  & n/a & 48.45$\pm$03.71 & 38.93$\pm$04.04 & 56.04$\pm$05.97 & 65.02$\pm$03.62 & \textbf{69.40$\pm$03.53} \\
    FBW    & n/a & 06.91$\pm$05.98 & 00.23$\pm$00.06 & 63.00$\pm$11.41 & 96.65$\pm$00.34 & \textbf{97.00$\pm$00.31} \\
    HepPh & n/a & 68.70$\pm$10.26 & 53.88$\pm$09.71 & n/a & 81.50$\pm$03.23 & \textbf{85.51$\pm$03.16} \\
    \midrule
          & \multicolumn{6}{c}{GR-AP@100-S1 (mean $\pm$ std)} \\
    \midrule
    AS733 & 01.31$\pm$00.31 & 88.87$\pm$04.08 & 95.39$\pm$01.68 & 83.15$\pm$06.52 & 96.50$\pm$00.51 & \textbf{97.15$\pm$00.49} \\
    Chess & 06.55$\pm$01.49 & 74.81$\pm$06.67 & 45.51$\pm$09.06 & 71.83$\pm$08.83 & 94.43$\pm$00.38 & \textbf{94.90$\pm$00.40} \\
    DNC   & 05.77$\pm$01.05 & 81.45$\pm$06.52 & 82.84$\pm$12.82 & 94.38$\pm$03.37 & 96.83$\pm$00.61 & \textbf{97.11$\pm$00.54} \\
    Elec  & n/a & 42.29$\pm$04.07 & 36.75$\pm$05.47 & 74.07$\pm$06.42 & 84.83$\pm$00.86 & \textbf{86.74$\pm$00.95} \\
    FBW    & n/a & 04.88$\pm$04.53 & 00.17$\pm$00.03 & 71.42$\pm$11.05 & 98.42$\pm$00.06 & \textbf{98.59$\pm$00.05} \\
    HepPh & n/a & 58.28$\pm$07.58 & 30.17$\pm$00.54 & n/a & 81.54$\pm$01.18 & \textbf{84.91$\pm$00.79} \\
    \midrule
          & \multicolumn{6}{c}{LP-AUC-S1 (mean $\pm$ std)} \\
    \midrule
    AS733 & 60.18$\pm$08.78 & 61.37$\pm$08.25 & 70.15$\pm$10.48 & 65.54$\pm$09.31 & 84.22$\pm$09.19 & \textbf{85.56$\pm$08.66} \\
    Chess & 64.23$\pm$08.63 & \textbf{86.66$\pm$10.00} & 85.77$\pm$12.46 & 79.32$\pm$10.29 & 77.83$\pm$05.60 & 77.24$\pm$05.87 \\
    DNC   & 75.90$\pm$08.83 & 84.18$\pm$10.39 & 89.34$\pm$12.40 & \textbf{90.30$\pm$07.75} & 79.18$\pm$06.05 & 78.76$\pm$07.96 \\
    Elec  & 66.32$\pm$11.22 & 91.16$\pm$06.99 & 82.83$\pm$15.31 & \textbf{97.31$\pm$02.32} & 89.38$\pm$03.86 & 90.31$\pm$03.81 \\
    FBW    & n/a & 82.83$\pm$10.58 & 82.88$\pm$17.82 & 81.76$\pm$07.10 & 87.39$\pm$01.32 & \textbf{88.00$\pm$01.19} \\
    HepPh & n/a & 88.39$\pm$10.36 & 82.37$\pm$12.35 & n/a & 88.52$\pm$08.75 & \textbf{90.25$\pm$07.61} \\
    \bottomrule
    \end{tabular}%
    }
  \label{Tab4}%
\end{table}%

\begin{table}[htbp]
  \centering
  \caption{CGR, GR and LP tasks under sliding way S2: the mean is given by the average over 20 time steps and over 10 runs; the std is given by the average of std over 10 runs; the best result is in bold.}
    \renewcommand\tabcolsep{6pt}
    \scalebox{0.92}{
    \begin{tabular}{l|cccccc}
    \toprule
          & GEM   & BCGD$^{l}$ & BCGD$^{g}$ & Triad & Ours$^{0.1,0.5}$ & Ours$^{0.2,0.5}$ \\
    \midrule
          & \multicolumn{6}{c}{CGR-AP@10-S2 (mean $\pm$ std)} \\
    \midrule
    AS733 & n/a & n/a & n/a & n/a & n/a & n/a \\
    Chess & 08.56$\pm$04.04 & 44.81$\pm$11.77 & 11.68$\pm$06.31 & 52.57$\pm$10.04 & 79.32$\pm$01.77 & \textbf{81.40$\pm$01.06} \\
    DNC   & n/a & 38.03$\pm$30.67 & 73.67$\pm$19.03 & \textbf{83.27$\pm$09.75} & 80.48$\pm$12.64 & 81.88$\pm$12.11 \\
    Elec  & 12.19$\pm$01.80 & 26.45$\pm$02.26 & 19.94$\pm$05.05 & 49.39$\pm$04.56 & 51.68$\pm$02.43 & \textbf{58.57$\pm$01.32} \\
    FBW    & n/a & 07.16$\pm$05.14 & 00.22$\pm$00.07 & 62.97$\pm$09.74 & 85.76$\pm$01.41 & \textbf{86.79$\pm$01.19} \\
    HepPh & n/a & 67.20$\pm$05.42 & 46.90$\pm$08.60 & n/a & 79.42$\pm$01.52 & \textbf{83.50$\pm$00.72} \\
    \midrule
          & \multicolumn{6}{c}{GR-AP@10-S2 (mean $\pm$ std)} \\
    \midrule
    AS733 & n/a & n/a & n/a & n/a & n/a & n/a \\
    Chess & 04.94$\pm$01.09 & 49.74$\pm$14.59 & 04.74$\pm$02.44 & 57.72$\pm$09.79 & 85.46$\pm$01.75 & \textbf{87.05$\pm$01.22} \\
    DNC   & n/a & 12.09$\pm$10.89 & 75.92$\pm$27.02 & \textbf{81.72$\pm$10.26} & 43.06$\pm$10.50 & 51.26$\pm$13.52 \\
    Elec  & 03.85$\pm$00.57 & 16.31$\pm$02.54 & 08.94$\pm$01.39 & 59.65$\pm$05.97 & 74.25$\pm$01.28 & \textbf{78.27$\pm$01.44} \\
    FBW    & n/a & 04.02$\pm$03.08 & 00.14$\pm$00.05 & 71.29$\pm$09.97 & 90.76$\pm$00.73 & \textbf{91.42$\pm$00.58} \\
    HepPh & n/a & 55.55$\pm$08.03 & 27.72$\pm$03.58 & n/a & 76.52$\pm$01.42 & \textbf{80.61$\pm$00.65} \\
    \midrule
          & \multicolumn{6}{c}{CGR-AP@100-S2 (mean $\pm$ std)} \\
    \midrule
    AS733 & n/a & n/a & n/a & n/a & n/a & n/a \\
    Chess & 11.89$\pm$04.41 & 71.83$\pm$07.75 & 51.93$\pm$18.80 & 67.12$\pm$05.60 & 89.92$\pm$04.60 & \textbf{90.66$\pm$04.31} \\
    DNC   & n/a & 73.01$\pm$23.73 & 79.92$\pm$21.92 & 95.18$\pm$04.54 & 99.23$\pm$01.35 & \textbf{99.31$\pm$01.15} \\
    Elec  & 12.47$\pm$02.14 & 39.33$\pm$05.81 & 33.54$\pm$14.63 & 61.68$\pm$03.42 & 60.92$\pm$04.13 & \textbf{66.88$\pm$02.59} \\
    FBW    & n/a & 11.45$\pm$07.43 & 02.50$\pm$05.86 & 76.47$\pm$07.73 & 98.55$\pm$00.69 & \textbf{98.68$\pm$00.64} \\
    HepPh & n/a & 62.21$\pm$02.81 & 45.87$\pm$08.75 & n/a & 80.37$\pm$01.40 & \textbf{84.51$\pm$00.48} \\
    \midrule
          & \multicolumn{6}{c}{GR-AP@100-S2 (mean $\pm$ std)} \\
    \midrule
    AS733 & n/a & n/a & n/a & n/a & n/a & n/a \\
    Chess & 08.91$\pm$02.05 & 82.25$\pm$04.74 & 59.78$\pm$18.36 & 74.99$\pm$05.97 & 96.53$\pm$01.38 & \textbf{96.79$\pm$01.28} \\
    DNC   & n/a & 82.51$\pm$15.20 & 88.54$\pm$18.57 & 98.42$\pm$01.84 & \textbf{99.85$\pm$00.34} & \textbf{99.85$\pm$00.33} \\
    Elec  & 04.55$\pm$00.79 & 40.10$\pm$06.30 & 41.57$\pm$12.62 & 76.17$\pm$04.01 & 83.99$\pm$00.92 & \textbf{86.05$\pm$00.63} \\
    FBW    & n/a & 08.60$\pm$07.52 & 02.69$\pm$06.54 & 83.52$\pm$07.36 & 99.34$\pm$00.33 & \textbf{99.39$\pm$00.30} \\
    HepPh & n/a & 55.84$\pm$05.18 & 30.96$\pm$05.92 & n/a & 82.55$\pm$01.28 & \textbf{85.83$\pm$00.63} \\
    \midrule
         & \multicolumn{6}{c}{LP-AUC-S2 (mean $\pm$ std)} \\
    \midrule
    AS733 & n/a & n/a & n/a & n/a & n/a & n/a \\
    Chess & 68.79$\pm$09.51 & \textbf{88.62$\pm$10.89} & 79.94$\pm$15.58 & 85.83$\pm$09.32 & 73.50$\pm$06.00 & 73.12$\pm$06.14 \\
    DNC   & n/a & 76.52$\pm$24.07 & \textbf{94.21$\pm$12.79} & 92.81$\pm$14.14 & 91.20$\pm$15.56 & 89.82$\pm$17.36 \\
    Elec  & 67.69$\pm$09.68 & 82.13$\pm$11.55 & 82.47$\pm$13.73 & \textbf{90.34$\pm$04.25} & 78.32$\pm$04.32 & 78.38$\pm$04.51 \\
    FBW    & n/a & 84.51$\pm$14.82 & 85.02$\pm$16.33 & 83.93$\pm$09.50 & 86.80$\pm$04.78 & \textbf{89.69$\pm$01.68} \\
    HepPh & n/a & \textbf{89.99$\pm$07.90} & 81.17$\pm$15.55 & n/a & 87.70$\pm$02.46 & 89.58$\pm$02.24 \\
    \bottomrule
    \end{tabular}%
    }
  \label{Tab5}%
\end{table}%

%We calculate std for the CGR, GR and LP of 21 snapshots in each run. Then we average the std in 10 runs to the average std.

Tables \ref{Tab4} and \ref{Tab5} show the more detailed results of the experiments in slicing ways S1 and S2 respectively. Differently form Table \ref{Tab3},  Table \ref{Tab4} and Table \ref{Tab5} additionally present the results of DynWalks$^{0.1,0.5}$ and the averaged standard deviation (std) i.e. the average over 10 runs of the std over 21 steps. According to tables \ref{Tab4} and \ref{Tab5}, we have the following observations.

For all tasks, DynWalks$^{0.2,0.5}$ achieves better results than DynWalks$^{0.1,0.5}$ in both S1 (see Table \ref{Tab4}) and S2 (see Table \ref{Tab5}) cases. But the performance gain is more significant in S2 case, which might be due to the changes of consecutive snapshots are larger than in S1 case. As a result, the more training samples may be required to reflect the larger changes, which leads to an interesting future work to make $\alpha$ adaptive based on the dynamic changes rather than choose a prefixed value. Note that, even if we compare DynWalks$^{0.1,0.5}$ (requires less computation resources than DynWalks$^{0.2,0.5}$) to other dynamic network methods, DynWalks$^{0.1,0.5}$ also receives the better results in most cases.

Regarding the stability w.r.t time-evolving, DynWalks is more stable than other methods in most cases. In particular, for CGR-AP@10 and GR-AP10 on DNC dataset, DynWalks does not receive the highest mean value, however, the std value of DynWalks is much smaller than the method that receives the highest mean value. And this observation is also according with the observation made in Figure \ref{Fig5}. Therefore, it enables DynWalks to work in some special real-world applications that require such stability w.r.t time-evolving.

\section{Global Topology and Recent Changes Awareness: The Removing Case}
\begin{figure}[htbp]
    \centering
    \includegraphics[width=0.9999\textwidth]{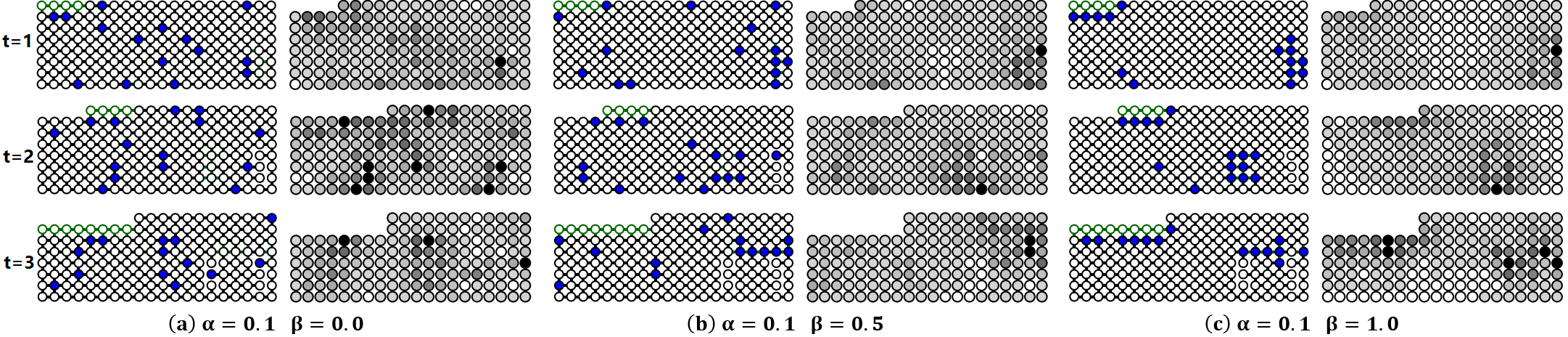}
    \caption{Balancing global topology and recent changes: the left ones of each sub-figure show the selected nodes in blue, and removed nodes/links are in green; the right ones show the number of times that nodes are trained in grey-scale. At $t=0$, the toy network is a regular 2d-grid with size (20, 8).}
    \label{Fig7}
\end{figure} 

In addition to Figure \ref{Fig4} i.e. nodes/links adding case, Figure \ref{Fig7} illustrates nodes/links removing case on a similar toy dynamic network. The general observations are similar as the observations made in Figure \ref{Fig4}. However, the choices of a proper $\beta$ might be more crucial for the removing case due to the following two reasons. Firstly, it may bring very sparsely connected part(s). However, random walks are naturally biased to densely connected part(s), and hence, may not generate enough node training pairs around the very sparsely connected part(s). And secondly, it may even bring isolated sub-network(s), so that random walks will be trapped inside the isolated sub-network(s) if without the help of enough diverse nodes.

\end{document}